\documentclass[journal]{vgtc}        

\ifpdf
 \pdfoutput=1\relax          
 \usepackage{graphicx}        
 \DeclareGraphicsExtensions{.pdf,.png,.jpg,.jpeg} 
\else
 \usepackage{graphicx}        
 \DeclareGraphicsExtensions{.eps}   
\fi%

\graphicspath{{figures/}{pictures/}{images/}{./}} 

\usepackage{microtype}         
\PassOptionsToPackage{warn}{textcomp} 
\usepackage{textcomp}         
\usepackage{mathptmx}         
\usepackage{times}           
\usepackage{cite}           
\usepackage{tabu}           
\usepackage{booktabs}         

\onlineid{0}

\vgtccategory{Research}
\vgtcpapertype{please specify}

\usepackage{xcolor}
\usepackage{float}
\usepackage{subfigure}
\usepackage{dblfloatfix} 
\usepackage{soul} 
\usepackage{enumitem}
\usepackage{afterpage}
\usepackage{float} 

\definecolor{cadmiumgreen}{rgb}{0.0, 0.42, 0.24}
\definecolor{darkyellow}{rgb}{0.72, 0.72, 0.15}

\newcommand{\ie}{i.e.,~}
\newcommand{\eg}{e.g.,~}

\newcommand{\red}[1]{\textcolor{red}{#1}}

\newcommand{\colorConfiguration}[1]{\textcolor{black}{#1}}

\newcommand{\rev}[1]{\textcolor{black}{#1}}
\newcommand{\revBar}[1]{\textcolor{black}{#1}}

\title{Multiscale Visual Drilldown for the Analysis of Large Ensembles of Multi-Body Protein Complexes}

\author{Katar\'ina Furmanov\'a, Adam Jur\v{c}\'{i}k, Barbora Kozl\'ikov\'a, Helwig Hauser, and Jan By\v{s}ka}

\authorfooter{
\item
K. Furmanov\'a, A. Jur\v{c}\'{i}k, and B. Kozl\'ikov\'a are with Masaryk University; E-mail: katarina.furmanova@gmail.com, \{xjurc,\,xkozlik\}@fi.muni.cz
\item 
H. Hauser is with the Univ.\ of Bergen; E-mail: Helwig.Hauser@UiB.no
\item 
J. By\v{s}ka is with the Univ.\ of Bergen and with Masaryk University; E-mail: jan.byska@gmail.com
}

\newcommand{\CD}{CE}
\newcommand{\CDFULL}{Complex Ensemble}

\newcommand{\CC}{CC}
\newcommand{\CCFULL}{Complex Configuration}
\newcommand{\CCFULLs}{Complex Configurations}
\newcommand{\PPC}{PPC}
\newcommand{\PPCFULL}{Protein Pair Configuration}
\newcommand{\PPCFULLs}{Protein Pair Configurations}
\newcommand{\PPD}{PPE}
\newcommand{\PPDFULL}{Protein Pair Ensemble}
\newcommand{\PPDFULLs}{Protein Pair Ensembles}
\newcommand{\AAP}{AAP}
\newcommand{\AAPFULL}{Amino Acid Pair}
\newcommand{\AAPFULLs}{Amino Acid Pairs}

\shortauthortitle{Furmanov\'a \MakeLowercase{\textit{et al.}}: TODO}

\abstract{
When studying multi-body protein complexes, biochemists use computational tools that can suggest hundreds or thousands of their possible spatial configurations. However, it is not feasible to experimentally verify more than only a very small subset of them.
In this paper, we propose a novel multiscale visual drilldown approach that was designed in tight collaboration with proteomic experts, \rev{enabling a systematic} exploration of the configuration space.
Our approach takes advantage of the hierarchical structure of the data -- from the whole ensemble of \colorConfiguration{protein complex configurations} to the individual configurations, their contact interfaces, and the interacting amino acids.
Our new solution is based on interactively linked 2D and 3D views for individual hierarchy levels and at each level, we offer a set of selection and filtering operations enabling the user to narrow down the number of configurations that need to be manually scrutinized.
Furthermore, we offer a dedicated filter interface, which provides the users with an overview of the applied filtering operations and enables them to examine their impact on the explored ensemble.
This way, we maintain the history of the exploration process and thus enable the user to return to an earlier point of the exploration.
We demonstrate the effectiveness of our approach on two case studies conducted by collaborating proteomic experts.

}


\CCScatlist{ 
 \CCScat{K.6.1}{Management of Computing and Information Systems}%
{Project and People Management}{Life Cycle};
 \CCScat{K.7.m}{The Computing Profession}{Miscellaneous}{Ethics}
}

\teaser{\vspace{2pt}
 \centering
 \includegraphics[width=1\linewidth]{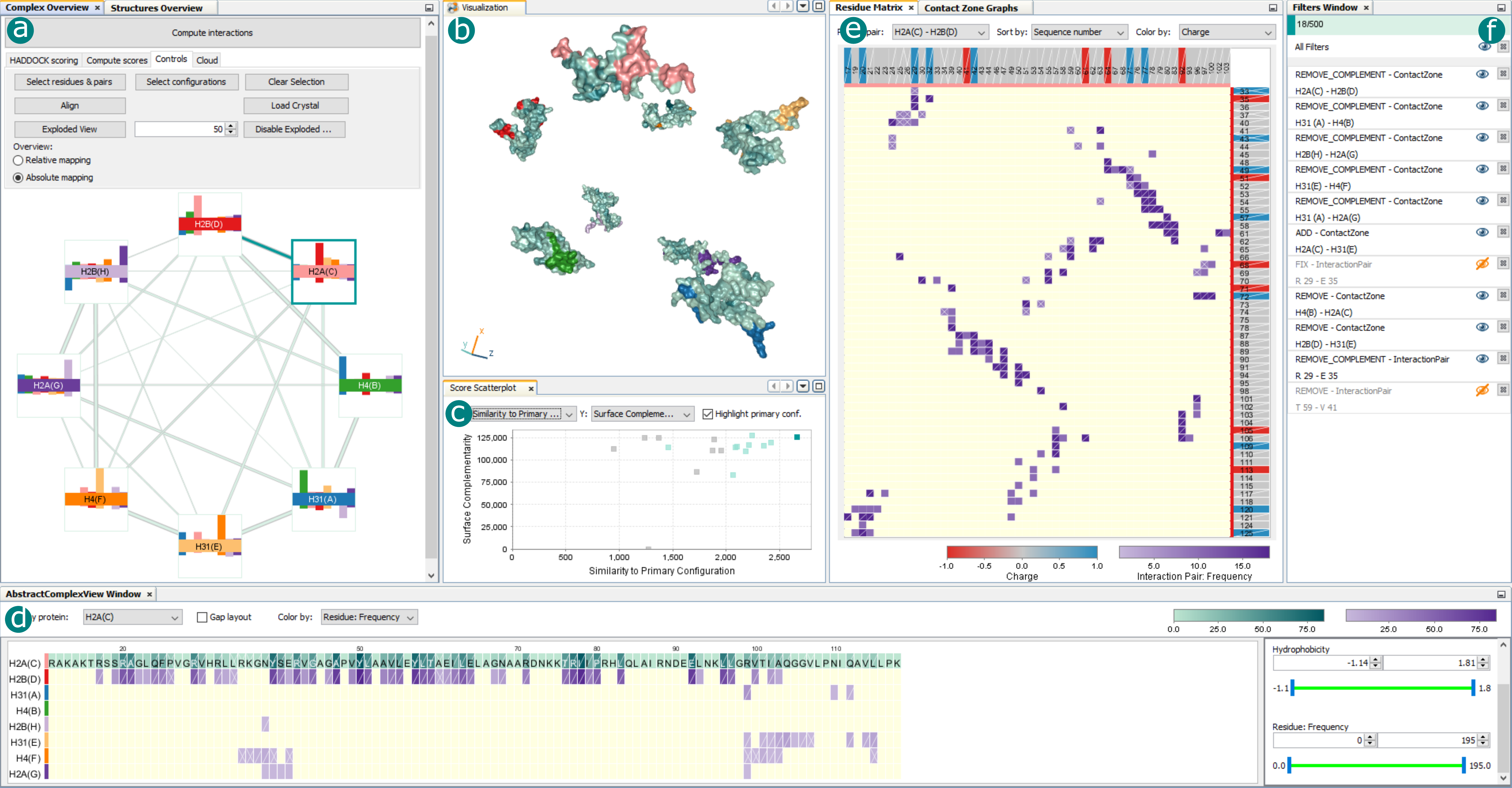}\\[-1.5ex] 
 \caption{Exploration of an ensemble of 500 possible human nucleosome configurations, based on the following components:~ 
 a)~Overview Graph with main application controls;~ 
 b)~3D View, showing the exploded \CCFULL, amino acids from contact interfaces are colored according to the frequency of their interactions;~ 
 c)~\rev{Property View}, showing properties of all \CCFULLs{};~ 
 d)~\rev{Protein View} with range filter panel;~ 
 e)~Residue Matrix, which also can be switched to Contact Zone List View,~ 
 and f)~Filter View.}
	\label{fig:teaser}
\vspace{5pt}
}

\vgtcinsertpkg

\begin{document}


\maketitle


\section{Introduction}

\begin{figure*}[tb]
	\centering
	\includegraphics[width=\linewidth]{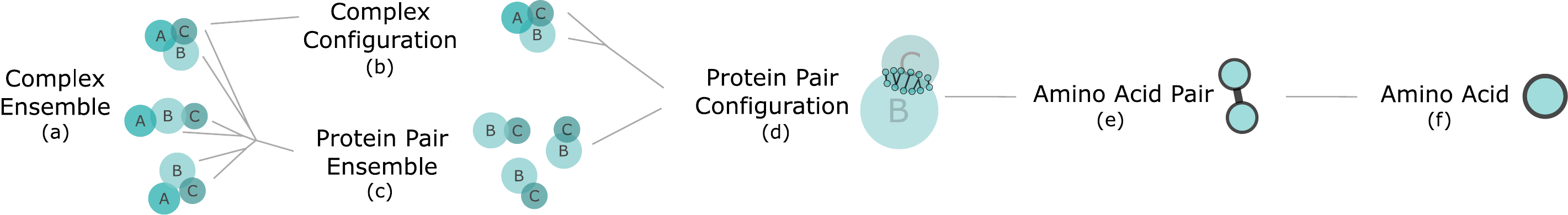}
	\vspace{-15pt}
	\caption{Structure of a protein complex ensemble containing possible spatial configurations of a protein complex consisting of proteins A, B, and C.}
	\vspace{-5pt}
	\label{fig:hierarchy}
\end{figure*} 

Protein complexes, formed by groups of associated polypeptide chains, facilitate a vast range of functions in living organisms, including the organization of DNA and metabolic processes. Furthering the understanding of protein functions is thus essential for advances in many areas of medicine, biology, and chemistry~\cite{palecek2019smc5, sledz2018protein}. 
The function of a protein complex is tightly bound to its spatial structure and the interactions between individual protein units. Revealing the structure of a protein complex requires costly and time-consuming experiments. 
Computational tools, predicting the mutual interactions of proteins, are therefore increasingly employed to aid experimental processes.

While most of the computational tools are limited to the prediction of binary complexes, many of the protein complexes consist of multiple units~\cite{levy20063d}.
To understand their function, it is necessary to take into account how all the units are interacting with each other. 
Irrespective of the computational approach, completely resolving the entire protein complex is a very costly combinatorial task, resulting in large ensembles of possible spatial arrangements of the protein complex.


Most of the multi-body docking tools, as well as several post-docking tools, also evaluate the resulting configurations with a scoring function, taking into account, e.g., shape complementarity at the inter-molecular interface, energies (Van der Waals, electrostatics, etc.), or knowledge-based potentials derived from known evolutionary process (e.g., conservation)~\cite{huang2008iterative, huang2014search}. The aim of the scoring function is to select the best representatives from the ensemble, containing thousands of docked configurations. Because of the large variety of the protein complexes and the differences between their structures and function, the tools and scoring functions are often tailored to a specific type of data and their usefulness for arbitrary protein complex is limited. \rev{This is also proven by ongoing CAPRI experiments~\cite{janin2003capri} that focus on the assessment of performance of the docking and scoring tools for newly identified protein complexes.}

As such, it is necessary for the domain experts to evaluate the results, assess the appropriateness of the scoring, and select the best representatives. Currently, this is done based on a cumbersome one-by-one exploration of the docked configurations in general molecular visualization tools such as PyMOL~\cite{PyMOL}, or in tools dedicated to the exploration of pairs of interacting proteins such as CoCoMaps~\cite{vangone2011cocomaps} or COZOID~\cite{furmanova2018cozoid}. These tools are tailored to the evaluation of binary docks, if at all, and lack a proper support for more intricate datasets. 

Therefore, we propose a novel interactive visual drilldown approach to the exploration of multi-body protein docking results. We take advantage of the hierarchical nature of the data and propose a system enabling domain experts to explore, compare, and filter protein complexes at different levels of detail. We propose several dedicated views, displaying the available information for each level in the data hierarchy. 
As the views are interactively linked, the users can observe how a given \rev{filter or selection} operation  translates to other levels of the data hierarchy. We also track all filtering operations and provide an interface where the users can check how previously used filters affect the explored ensemble and, if necessary, revert them. 

\rev{The bi-directional linking across several levels of such hierarchy is a challenging task as the propagation of the user's actions through the levels of the hierarchy can lead to large changes in the individual views. As this can be confusing, we have to carefully consider how and when the user-triggered changes are propagated. As hierarchical data appear in many domains, the presented solution for multi-level filtering and selection is one of the main contributions of this paper.}

\section{Tasks and Data Analysis}
\label{sec:data_tasks}
\rev{
Our solution was designed in tight collaboration with proteomic experts from two research groups, one focusing on structural biology and one on protein engineering.
We conducted several focus groups at the beginning of the design process as well as after the experts had a chance to test the first prototype. 
The main topic of the focus groups was the discussion of the current workflow and the identification of challenges connected to the analysis of multi-body protein complexes.}

\rev{In the initial focus group we discovered that} proteomic experts explore results from computational docking tools to identify the biochemically most relevant spatial arrangements of the examined protein complex that can be experimentally verified in the lab.
When performing this analysis, the domain scientists draw heavily on existing knowledge from the literature. 
To be able to incorporate this knowledge during the analysis process, the proteomic experts need to be able to quickly explore and filter the predicted arrangements \rev{concerning} a multitude of biochemical and spatial properties on different levels of detail.
Without a proper visual support, this is a very tedious and time-consuming process as they need to explore large and intricate ensembles of predicted configurations.


We identified the following hierarchical structure of the data for protein complexes. We call the initial dataset produced by a computational docking tool a \textbf{\CDFULL{}} \rev{(\CD{}, Figure~\ref{fig:hierarchy}a)}, containing the entire dataset of all computed configurations. It consists of many \textbf{\CCFULLs{}} (\CC{}, \rev{Figure~\ref{fig:hierarchy}b}), where each configuration represents one possible spatial arrangement of the whole protein complex.
The \CCFULL{}
can be further split into individual \textbf{\PPCFULLs{}} (\PPC{}, \rev{Figure~\ref{fig:hierarchy}d}), each representing exactly one mutual position of two proteins 
from the protein complex. Each \PPCFULL{}
contains a set of \textbf{\AAPFULLs{}} (\AAP{}, \rev{Figure~\ref{fig:hierarchy}e}), each consisting of two interacting \textbf{amino acids} (AA, \rev{Figure~\ref{fig:hierarchy}f}) from the two proteins 
forming the \PPCFULL{}.
\rev{Alternatively, we can split the \CDFULL{} into different \textbf{\PPDFULL{}s}} (\PPD{}, \rev{Figure~\ref{fig:hierarchy}c}), each 
consisting of the set of all \PPC{}s that are formed by the same protein pair.
\rev{In our solution, we employ run-time data structures that correspond one-to-one to the above-described structure.
Such structure was sufficient to perform filtering and selections in real time across all scales in all tested datasets.}


At each level of this hierarchy, we observe different properties, which can be explored and used for the identification of the biochemically most relevant \CC{}s. 
\rev{Individual \CC{}s and their \PPC{}s can be scored based on energy measures between the interacting atoms (\eg Van der Waals or electrostatic energies), where smaller numbers indicate more favourable interactions, or based on their geometrical properties (\eg surface complementarity).}


The docking tools often use a combination of these and other properties to define their own scores and select the most relevant representatives. However, these scoring functions \rev{can give different results for} different protein complex types. Thus, \rev{the experts cannot fully rely on this ranking and filter out some solutions solely based on the scores of the computational tools.} 

In our solution, these properties \rev{and scores} can be pre-loaded from the docking results, if provided (e.g., from HADDOCK~\cite{karaca2010haddock_multibody}). 
Additionally, we include a property indicating the similarity of \CCFULLs{} (\CC{}) and \PPCFULLs{} (\PPC{}) to one selected, so-called primary \CC{} or \PPC{}. This \rev{measure} is useful in cases when the domain expert already identified a potentially relevant \CCFULL{} and searches for similar \CCFULLs{}, possibly revealing an even better solution. We also enable to compute this similarity \rev{concerning} a protein complex that is not part of the explored ensemble, to include the domain knowledge from partially resolved models or to compare results among different species. The similarity score computation is adapted from our previous work~\cite{furmanova2018cozoid}. 

\rev{In terms of interacting amino acids, the domain experts are often interested in exploring the frequency of their occurrence in a dataset, distances between them, or their physico-chemical properties, such as hydrophobicity (\ie tendency to interact with water molecules) or charge, both indicating the feasibility of a given interaction.}



\rev{Based on the focus groups with domain experts,} we identified a set of tasks that a visualization system for the analysis of large ensembles of multi-body protein complexes should support:

\newcommand{\TGOAL}{\textbf{T0}}
\newcommand{\TEXPPCC}{\red{REPLACE WITH TPROT}}
\newcommand{\TPROT}{\textbf{T1}}
\newcommand{\TCZ}{\textbf{T2}}
\newcommand{\TAA}{\textbf{T3}}
\newcommand{\TFILTER}{\textbf{T4}}
\newcommand{\TCOMPARE}{\textbf{T5}}
\newcommand{\TPROV}{\textbf{T6}}

\begin{itemize}[noitemsep,nolistsep]
  \item[\TGOAL{}]The primary goal is to identify the biochemically most relevant \CCFULLs{} from a large ensemble of predicted spatial arrangements.
  \item[\TPROT{}]To enable this, we need to provide the ability to explore all \CCFULLs{} and their spatial and biochemical properties, and support the identification of the interacting proteins.
  \item[\TCZ{}]It is also necessary to provide means for quick identification of potentially important \PPDFULLs{} and the exploration of individual \PPCFULLs{} and their properties.
  \item[\TAA{}]The solution has to enable the domain experts to identify the interactive \AAPFULLs{} and explore their properties both globally and separately for each \PPCFULL{}.
	\item[\TFILTER{}]In order to decrease the size of the exploration space, it is necessary to filter out invalid predictions based on their properties on all levels of the hierarchy as soon as they are discovered.
	\item[\TCOMPARE{}]To exploit domain knowledge and ease the exploration, experts require the ability to compare the main differences between individual arrangements as well as those when comparing to other protein complexes.
	\item[\TPROV{}]Furthermore, the system should provide information about the exploration steps that the domain experts undertook and enable the expert to adjust or revert them. 
\end{itemize}
\section{Related Work}
\label{sec:relatedwork}
Our work is related to several research areas which we can generally divide into two main parts.
The first part is dealing with computational approaches, producing ensembles of possible configurations.
The second area covers the visual exploration of these ensembles, including the visual representation of protein complexes on different levels of detail and their visual analysis using drilldown and filtering.

\subsection{Computation of Protein Complexes}
Most of the currently available computational tools for protein-protein interactions are focusing on protein pairs and a comprehensible overview was published by Huang~\cite{huang2014search}.
Some of the existing approaches, such as ArDock~ \cite{ardock}, already combine the computational method with a basic visual representation of the predictions. 
There are even some solutions, such as DockingShop~\cite{dockingshop}, which are enabling the user to interactively design an initial configuration for a protein docking prediction process through a molecular graphics interface.

In the past decade, several tools emerged which enable the prediction of multi-body protein complexes.
To compute a spatial arrangement of bigger assemblies, some computational tools reuse the pair-wise docking of the involved proteins and combine them once computed (\eg\cite{esquivel2012lzerd, venkatraman2012antcolony}), while others rely on restraints derived from experimental data to limit the exploration space (\eg\cite{karaca2010haddock_multibody, kuzu2016prism}). 

One of the first tools designed primarily for multi-body docking was CombDock~\cite{inbar2005prediction}. 
The algorithm works on a principle of hierarchical construction of the complex from smaller subunits and a greedy selection of the best-ranking subunits. 
The combinatorial step is followed by the reduction of solutions based on RMSD and a scoring function. 
Multi--LZerD~\cite{esquivel2012lzerd} uses a genetic algorithm to generate complexes from initial pairwise docks and applies an energy minimization structure refinement procedure for the ranking of the solutions. 
Venkatraman et al.~\cite{venkatraman2012antcolony} proposed an ant colony optimization approach to solve the combinatorial problem. DockStar~\cite{amir2015dockstar} formulates the task of detecting the spatial conformation of a protein complex as an Integer Linear Program. 
Unlike other methods, it also integrates experimental data from mass spectrometry into the scoring of the solutions. 
Another tool reusing pairwise docks in combination with experimental data is PRISM-EM~\cite{kuzu2016prism}. 
It uses density maps from cryo-electron microscopy for guiding the placement of subunits. 

While all of the tools mentioned above build the complex structure from samples of pairwise prediction, HADDOCK~\cite{karaca2010haddock_multibody}, to the best of our knowledge, is the only tool that performs simultaneous docking of multi-body complexes. 
During the docking process, it uses constraints from different kinds of experimental data to drive the formation of the complex. 
However, unlike DockStar~\cite{amir2015dockstar}, which supports the prediction of complexes consisting of up to 16 units, HADDOCK
allows predicting only complexes with up to 6 units.

\subsection{Visualization of Protein Complexes}
Protein complexes, irrespective of the number of units, can be in general visualized in one of the traditional and widely used software tools, such as UCSF Chimera~\cite{Chimera}, PyMol~\cite{PyMOL}, VMD~\cite{vmd}, Aquaria~\cite{aquaria}, CAVER Analyst~\cite{caverAnalyst}, and others.
However, these tools provide the users only with general molecular representations, without specific support for the analysis of protein complexes and their visual exploration.
This need was addressed by Lee and Varshney~\cite{Lee2006} by visualizing the volume of the area of the docking site between two proteins.
Another attempt of a schematic visualization of two interacting proteins was proposed in the DIMPLOT extension of the LigPlot+ tool~\cite{ligplot}.
DIMPLOT shows the protein-protein interface in a 2D diagram.
The 3D representation of this diagram can be viewed in PyMOL~\cite{PyMOL} or RasMol~\cite{rasmol,rasmolab}.
A 2D schematic representation of interacting areas between proteins is available also in the PDBsum database~\cite{pdbsum}.

Although these schematic representations
are conveying the information about a single configuration, they do not support the comparison and interactive filtering of entire ensembles of configurations.
This issue is addressed in the CoCoMaps~\cite{vangone2011cocomaps} and COZOID~\cite{furmanova2018cozoid} tools.
Both tools come with linked visualizations, aiding the users in analyzing and comparing interactions between protein pairs.
CoCoMaps and its successor CONS-COCOMAPS~\cite{conscocomaps} enable to measure and visualize the consensus in multiple docking solutions and display the conservation of residue contacts using intermolecular contact maps.
The COZOID tool uses a set of linked views for the interactive exploration of large ensembles of protein pairs, supporting a visual drilldown approach for narrowing down the set of possibly relevant configurations.
The main limitation of these approaches is that they are operating only on protein pairs (i.e., single \PPDFULL{}) and cannot be directly applied to multi-body complexes.
The multiscale aspect in molecular visualization can be explored on different granularity levels, as shown in the recent survey of Miao et al.~\cite{Miao2019}.

In our case, we were not only concerned with designing proper visual representations of the individual hierarchy levels of large ensembles of multi-body complexes, but also with how to interactively explore and filter these ensembles to support the identification of biochemically most relevant instances.
Splechtna et al.~\cite{Splechtna2015} focus on the problem of interactive visual steering of hierarchical simulation ensembles.
In their substantially different application case, they also deal with linking representations on different levels of detail as well as with the challenge that the ensemble can grow during the exploration process.


\section{Method Overview}
\label{sec:method}
The primary goal of this work was to enable biochemists to identify the biochemically most relevant spatial arrangements of multiple interacting proteins (\TGOAL{}) from a large ensemble of possible \CCFULLs{} (\CC{}), predicted by a computational tool. 
\rev{To support this goal, we designed our solution as an interplay between several linked 2D and 3D views. Figure~\ref{fig:teaser} shows an overview of the solution and all its components. The high-level overview of all loaded \CC{}s is available in the Overview Graph (a) which serves for the preliminary filtering of \CC{}s for further scrutinization. Details about selected properties of \CC{}s of interest are visible in abstract views (c), (d), and (e), and the most detailed view of individual \CC{}s is the 3D view. Here we provide an exploded view as the most interesting parts of the complex are the contact zones, hidden inside the complex. Due to the option of advanced filtering operations, we also added a specific panel supporting this task. In the following, we first describe the proposed interaction concept and then the individual views in more detail.}

\subsection{Interaction Design}
\label{sec:intearction_design}

\rev{The phenomena at hand have a natural hierarchical structure where all levels of hierarchy have to be explored (\TPROT{}, \TCZ{}, \TAA{}) by domain scientists.}
The users can perform filtering and selection operations in each view to support the exploration of a large number of \CCFULLs{} (\CC{}). The filtering operations define rules describing which \CC{}s are irrelevant and thus should not be displayed to the user in subsequent analysis steps (unless the user specifically recalls them to change/reset the filtering) it is also interactively determined which \CC{}s are important and should be kept. The selection operations, on the other hand, define the current focus of the user, allowing to highlight more important data while suppressing the context. 

\smallskip
\vspace{5pt}
\noindent We elicited four primary filtering operations: 
\begin{itemize}[noitemsep, nolistsep]
  \item \textbf{Remove:} filters all \CCFULLs{} satisfying a rule, i.e., all selected \CC{}s or all \CC{}s with a selected \AAPFULL{}.
  \item \textbf{Remove complement:} filters all \CCFULLs{} not satisfying a rule.
  \item \textbf{Fix:} prevents \CCFULLs{} from being removed from the ensemble by other filtering operations -- this operation does not remove anything.
  \item \textbf{Add:} re-adds the previously filtered \CCFULLs{} back to the ensemble for further exploration.
\end{itemize}
\smallskip
When designing the filtering operations, we have considered merging the \textit{remove complement} and \textit{fix} operations into one. An exemplary case for such a merged operation would be the following: the user knows from experimental results that a certain pair of amino acids must interact in the protein complex. Thus, he or she would search for the given \AAPFULL{} and fix it and remove all \CCFULLs{} that do not contain this pair. However, this would effectively block the user from any further drilldown, since all the \CCFULLs{} containing this \AAPFULL{} (potentially still a large number) would be fixed. Therefore, we decided to separate the operations. While the \textit{remove complement} operation is useful in the above-described example, the \textit{fix} operation is useful, e.g., in cases, where the user identifies potentially interesting \CCFULL{}s and does not want to lose them during the continued drilldown process, but wishes to explore other \CCFULLs{}.

We have also considered the usefulness of the \textit{add} operation as the same effect can be achieved by reverting filters or by using the \textit{fix} operation. However, reverting filters can be impractical, since it can also re-add unwanted data items back to the ensemble and the desired items can also be affected by multiple filters. Fixing the items is also not a solution when the user wants to take a closer look at the \CCFULLs{}, but is not yet sure if they are biochemically relevant.

Our filtering operations are based on the selection of single or multiple elements. Dedicated views enable the selection of the elements at different levels of the hierarchy. As we have explained in Section~\ref{sec:data_tasks}, data items can be equipped with multiple properties, that can serve as a basis for filtering. Therefore, we have also introduced range filters that operate on the quantitative properties of the data items. Range filters can be considered as a special type of \textit{remove complement} filters, where the items 
not satisfying the given range condition are removed.

Additionally, a dedicated Filter View offers the users an overview of the currently set filtering operations. Here the users can choose to revert a filter or temporarily disable some filters to see the items they affect and possibly re-add them back to the ensemble.
Filters are evaluated in the order in which they were added. Exceptions to this are the \textit{fix} operations, which override all other filters.
The position of range filters in the evaluation queue is updated each time the range is changed.

Both filtering and selection operations can be performed on any level of detail. Therefore, we had to consider how to propagate the operations through the data hierarchy and the linked views. This propagation clearly needs to be bi-directional, meaning that if the operation is performed at a lower level of the hierarchy, it is first propagated to the top level and then to the lower levels again. For example, if the user removes an \AAPFULL{} (\AAP{}), all the \CCFULLs{} containing this \AAP{} will be identified and filtered from the ensemble. Consequently, all their \AAP{}s and interacting amino acids will be removed as well. \rev{Therefore,} removing one \AAP{} in a particular view likely results in filtering other \AAP{}s from the same view. 

Similarly, if the selection propagation was automatic, the selection of one \AAP{} in a particular view would lead to the selection of all the \AAP{}s of all the \CCFULLs{} containing the initially selected \AAP{}. However, unlike with filtering, where the filtering of items upon setting a filter is expected, the automatic selection propagation is difficult to control and understand. Therefore, we decided to make the selection propagation between different levels of data hierarchy on demand. 

 
\subsection{Overview Graph}
\label{sec:overview}
\begin{figure}[tb]
	\centering
	\includegraphics[width=0.80\columnwidth]{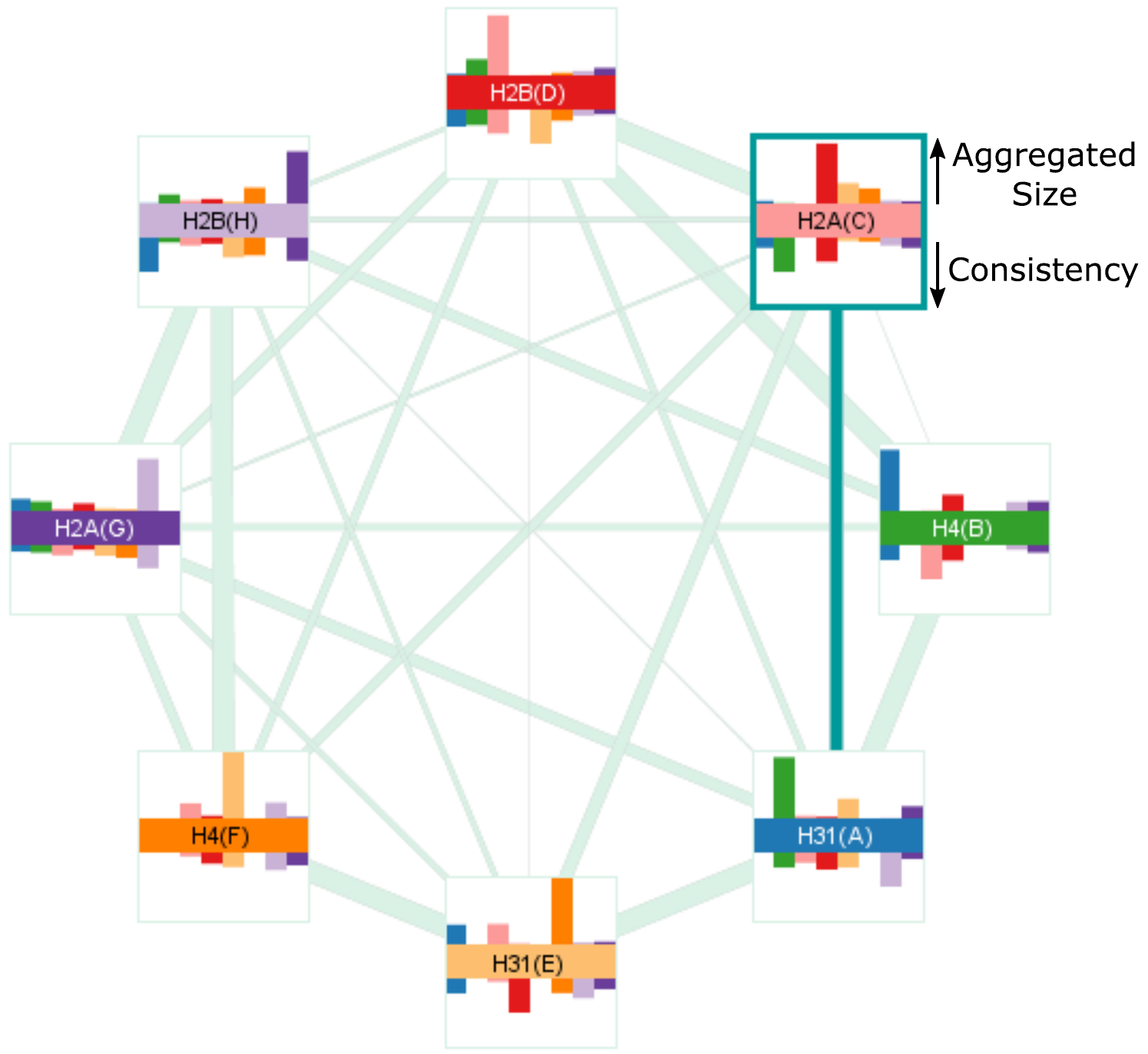}
	\caption{Overview Graph showing a protein complex with of 8 interacting proteins. The primary \PPDFULL{}, related to H2A(C) and H31(A), is highlighted in darker color, as well as the primary protein H2A(C).}
	\vspace{-5pt}
	\label{fig:overview}
\end{figure}

The Overview Graph (Figure~\ref{fig:overview}) was designed to provide summary information about the entire ensemble (\TPROT{}) and to aid the navigation during the exploration process. It represents the ensemble in the form of a node-link diagram, where nodes correspond to the individual proteins of the complex, while edges between nodes represent the presence of an interaction between the corresponding proteins. We opted for a simple circular layout of the graph as it is easily readable for a small number of nodes. For protein complexes, this is sufficient as the number of proteins in a complex rarely exceeds ten.

To support the navigation to the views depicting the lower levels of the data hierarchy, which are either tied to a single protein or a pair of proteins (i.e., \PPDFULL{} \rev{(\PPD{})}), \rev{the users can select a \textit{primary protein} (node) and a \textit{primary \rev{\PPD{}}} (edge) in the Overview Graph. These are then highlighted with darker color (see Figure~\ref{fig:overview}).}


Each graph node consists of \revBar{three} parts. The central part of each node is colored by the color assigned to the represented protein from a predefined scheme ensuring the colors are well distinguishable. This color is then used consistently in other views to identify individual proteins. Each node is surrounded by \revBar{two bar charts} depicting the aggregated sizes \revBar{and the consistency} of the contact interfaces between the interacting proteins.

\revBar{The information about the sizes of the interfaces is mapped onto the upper bar chart. The height of each bar is computed as a sum of the \AAPFULLs{} that are present in the respective \PPDFULL{}. The same \AAPFULL{} appearing in multiple \PPCFULLs{} is counted just once.}
\revBar{The users can choose to either scale the bar charts independently for each protein such that the bar representing the largest interface always fills the available space or scale them using absolute values. While the former approach is better utilizing the available space, the latter allows comparing the sizes of the interfaces across the whole protein complex. However, it is less} suitable for complexes where proteins vary in size -- and consequently in the sizes of their contact interfaces -- as the representative \revBar{bars} for small but still important contact interfaces \rev{may} become too small.




\renewcommand{\thefigure}{5} 
 \begin{figure*}[b] 
	\centering
	\includegraphics[width=\linewidth]{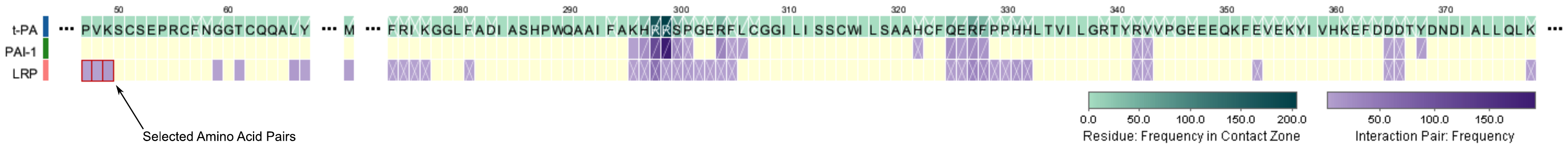}
	\caption{\rev{Protein View} of t-PA protein (primary) with the selected \AAPFULLs{} interacting between LRP protein and the Finger domain of t-PA. The AAs and \AAPFULLs{} that were filtered out based on this selection are indicated with white crosses and diagonals.}
	\label{fig:casestudy-1D}
\end{figure*}

Usually a large \revBar{bar} corresponds to a large contact interface between the proteins. However, since \rev{the \revBar{bars}} represent the aggregated information, the \revBar{bar} for a \PPDFULL{} with a small but varying set of contact interfaces can be equal to the \revBar{bar} for a \PPDFULL{} with large and consistent contact interfaces. Therefore, we indicate the \textit{consistency} of the contact interfaces \revBar{in the bottom bar chart} -- the \revBar{bigger bar signifies that the \AAPFULLs{} (\AAP{}) in the contact interface are more consistent.} The \textit{consistency} is computed as $\frac{1}{N_{\AAP{}}} \sum_{\small{\AAP{}}} {P_{\small{\AAP{}}}}$, where $N_{\AAP{}}$ is the number of unique \AAP{}s in the contact interface (i.e., the value \revBar{depicted in the upper bar chart}) and $P_{\AAP{}}$ is the percentage of the \PPCFULLs{} (\PPC{}) from all \PPC{}s where the given pair is present. The light \revBar{border} of the \revBar{bar chart} provides a reference value \revBar{both} for the \revBar{maximum size of the interface as well as} for the \textit{consistency} equal to 1, i.e., the cases where all \PPC{}s have exactly the same contact interface. 


The width of the links between proteins encodes the number of \CCFULL{}s where these two proteins interact, i.e., the size of the \PPDFULL{} corresponding to these proteins. The users can interact with the edges via a pop-up menu to select or filter all \CCFULLs{} containing (or not containing) the contact between the corresponding proteins. 

\subsection{\rev{Property View}}
\label{sec:scatterplot}
As we have described in Section~\ref{sec:data_tasks}, \CCFULLs{} (\CC{}) can be evaluated according to various physico-chemical properties, such as the energetic score or surface complementarity. As these properties are often used for scoring and prefiltering of the configuration space, it is important for users to understand the relationship between the scoring and the dataset of \CCFULLs{} -- i.e., the distribution of \CC{}s according to various properties and the correlation between individual properties (\TPROT{}, \TCOMPARE{}). In the \rev{Property View} (see Figure~\ref{fig:scatterplot}), each \CC{} is depicted as a point with coordinates set to two user-selected properties \rev{(i.e, we use a scatterplot representation)}. Users can select the \CC{}s using rectangular brushing or clicking on individual \CC{}s. 

\renewcommand{\thefigure}{4} 
\begin{figure}[h]
	\centering
	\includegraphics[width=\columnwidth]{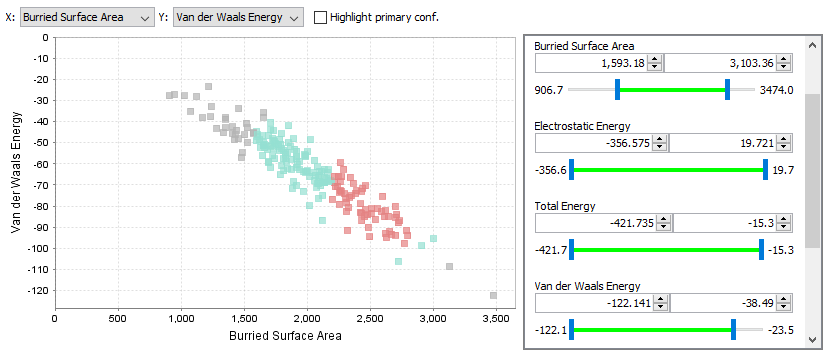}
	\caption{\rev{Property View} with the filtering panel. Selected \CCFULLs{} are shown in red, \CC{}s unaffected by filters are depicted in green, and \CC{}s filtered out by a temporarily disabled filter are shown as grey.}
	\label{fig:scatterplot}
	\vspace{-5pt}
\end{figure} 

\subsection{\rev{Protein View}}
\label{sec:1Dview}



In \rev{Protein View} (see~Figure~\ref{fig:teaser}d), we provide the users with more detailed information about the interaction of the selected primary protein with other proteins in the {\CDFULL{}}.
It shows the individual amino acids \rev{(AA)} of the primary protein \rev{in the first row} and encodes how often these interact with other proteins. 
Each \rev{other} row corresponds to one protein in the complex that interacts with the primary protein in at least one \CCFULL{}. The cells of the row then show interactions from \PPDFULL{} of \rev{the} primary protein and the protein corresponding to the given row. This \rev{layout} enables the domain experts to identify potentially important \PPDFULLs{} that should be further explored (\TCZ{}). \rev{Individual rows are equipped with the name and color bar identifying individual proteins.}

We map the information about the interaction frequency of AAs to the color -- green for the overall number of interactions of the amino acids from the primary protein and purple for the number of interactions of these amino acids with individual proteins in other rows. The depicted frequency is always derived from the visible (i.e., unfiltered) portion of the ensemble. The coloring of the AAs of the primary protein can be changed to reflect other AA properties.

This view enables domain experts to immediately see which parts of the primary protein are interacting with other proteins (\TAA{}) and how often. It also provides means to identify amino acids from the primary protein that are potentially interacting with multiple proteins. In nature, these are important but rare cases, as such bonding only happens under specific conditions (e.g., specific spatial orientation of the amino acids). In docking predictions, it is more often the case of an incorrect bonding. The proteomic experts are, therefore, interested in locating and verifying such interactions.

The user can choose to depict either all amino acids from the primary protein or to use a condensed view (see~Figure~\ref{fig:casestudy-1D}).
\rev{In the condensed view, we hide the amino acid sequences longer than 25 amino acids that do not have any interaction.}
This value was experimentally chosen to remove long and uninteresting sequences from the view but also to avoid too much fracturing of the sequence, which would decrease the readability of the view as there are usually many smaller gaps between interacting amino acids. The ruler on top of the view indicates the sequence number of every 10th \rev{amino acid} for improved orientation.

The \rev{view} supports various filters, allowing users to filter the \CCFULLs{} based on the properties of individual amino acids from the primary protein (using range filters) as well as to explicitly enforce the presence (or absence) of the selected amino acids.
Additionally, selecting a single row will result in setting the interface between the primary protein and protein represented by the selected row as the primary \PPDFULL{} that can be later explored in other views. Selection of cells will highlight the corresponding \AAPFULLs{} and AAs in other views, including the AA positions in 3D.


\subsection{Residue Matrix}
\label{sec:residuematrix}
The \rev{Protein View} allows to explore the interacting amino acids of one selected protein. However, it does not provide any detailed information about its counterparts. Therefore, to fully address task \TAA{}, we have adapted the solution which we designed in our previous tool~\cite{furmanova2018cozoid} for the visual analysis of protein interaction pairs. When the user decides which \PPDFULL{} (\PPD{}) to explore, he or she can get an aggregated overview of all \AAPFULLs{} in the selected \PPD{} using the Residue Matrix (see Figure~\ref{fig:teaser}e).

Similarly to the \rev{Protein View}, the axes of the Residue Matrix show the interacting amino acids from two proteins corresponding to the selected \PPD{}. For easier navigation, the proteins are identified by the colored lines along the matrix axes. The cells in the matrix indicate the frequency of the occurrences of the \AAPFULL{} in the \PPD{} and are also derived only from the visible portion of the ensemble. 
We extended the original view~\cite{furmanova2018cozoid} with new options for sorting and coloring of the AAs forming the matrix axes, e.g., using properties such as charge and hydrophobicity in addition to the interaction frequency. 

We further extended the original view with the support for multiple filters. The users can easily select a valid range of various biochemical properties for both individual amino acids and \AAPFULLs{}. We also allow filtering of the \CCFULL{} based on the explicit manual selection of \AAPFULLs{} that need to (or cannot) be present in the \PPD{} to be biochemically relevant.


\subsection{Contact Zone List View} 
\label{sec:cozolists}
To address task \TCOMPARE{}, we have adapted another solution which we proposed in the COZOID tool~\cite{furmanova2018cozoid}, the Contact Zone List View (see Figure~\ref{fig:cozolist}).
This view depicts individually selected \PPCFULLs{} (\PPC{}) side-by-side. Each \PPC{} is represented by two columns of amino acids coming from the two proteins forming the primary \PPDFULL{}. \rev{This view aims} 
to offer a detailed \rev{representation} of a small number of selected \PPC{}s.
\rev{The \PPC{}s are by default ordered by their similarity to the reference \PPC{} (which can be one of the \PPC{}s in the explored ensemble or a separate \PPC{}, e.g., of a partially resolved crystal from a protein database).} 
The view also offers the comparison mode in which the main similarities (and differences) between the reference \PPC{} and the rest are highlighted by the presence (and absence) of colored borders, edges, or even the whole amino acid boxes. Similarly to the Residue Matrix, we have extended this view with new options for sorting the \PPC{}s \rev{based on the summary properties} as well as sorting and coloring of the AAs in the lists.


\renewcommand{\thefigure}{\arabic{figure}} 
\begin{figure}[tb]
	\centering
	\includegraphics[width=1\linewidth]{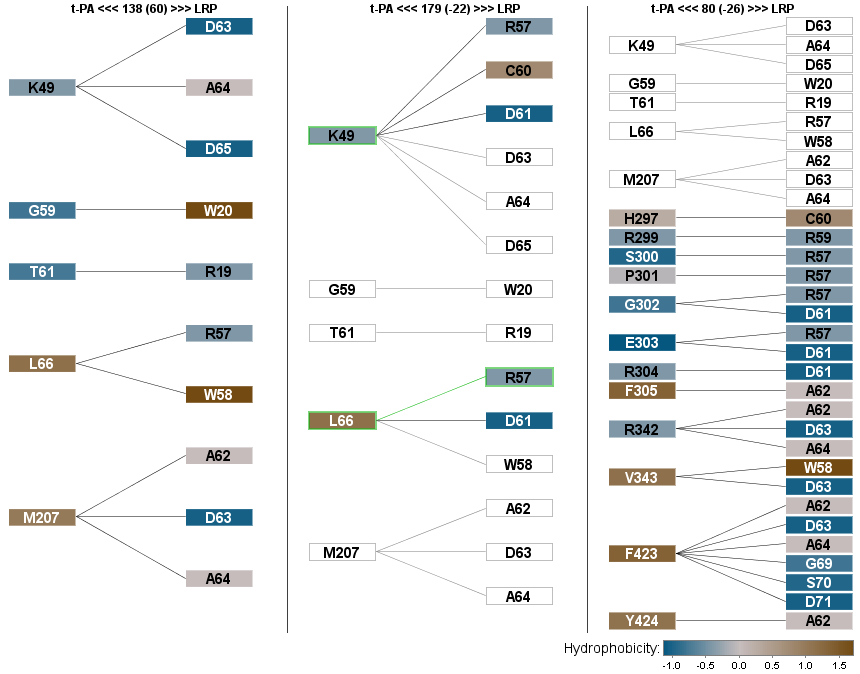}
	\caption{Contact Zone List View showing the comparison of \AAPFULLs~from the primary \PPCFULL{} (\PPC) (left) with two other \PPC{}s (middle, right). AAs and \AAPFULLs{} of compared \PPC{}s corresponding to the primary \PPC{} are highlighted in green, while elements present in primary \PPC{} but missing in compared \PPC{}s are shown with empty cells. It can be seen that only the first of the compared \PPC{}s shares some amino acids (K49, L66, R57) with the primary \PPC{}.}
	\label{fig:cozolist}
\end{figure} 

\subsection{3D Views}
\label{sec:3Dview}

To depict the actual spatial arrangement of individual \CCFULL{}s, which is part of task \TPROT{}, we utilize a 3D view. Our system supports all standard molecular representations 
which are, however, either too abstract or they suffer from occlusion and hide the most interesting parts of the protein complex -- the contact interfaces between proteins. Moreover, the standard views are unusable for the exploration of larger ensembles of \CCFULLs{} (\CC{}s), as showing multiple \CCFULLs{} at the same time results in high visual clutter and occlusion. Users can browse the individual \CC{}s one-by-one, however, for an ensemble containing hundreds of \CC{}s this approach is infeasible. 

\begin{figure}[tb]
	\centering
	\includegraphics[width=0.6\columnwidth]{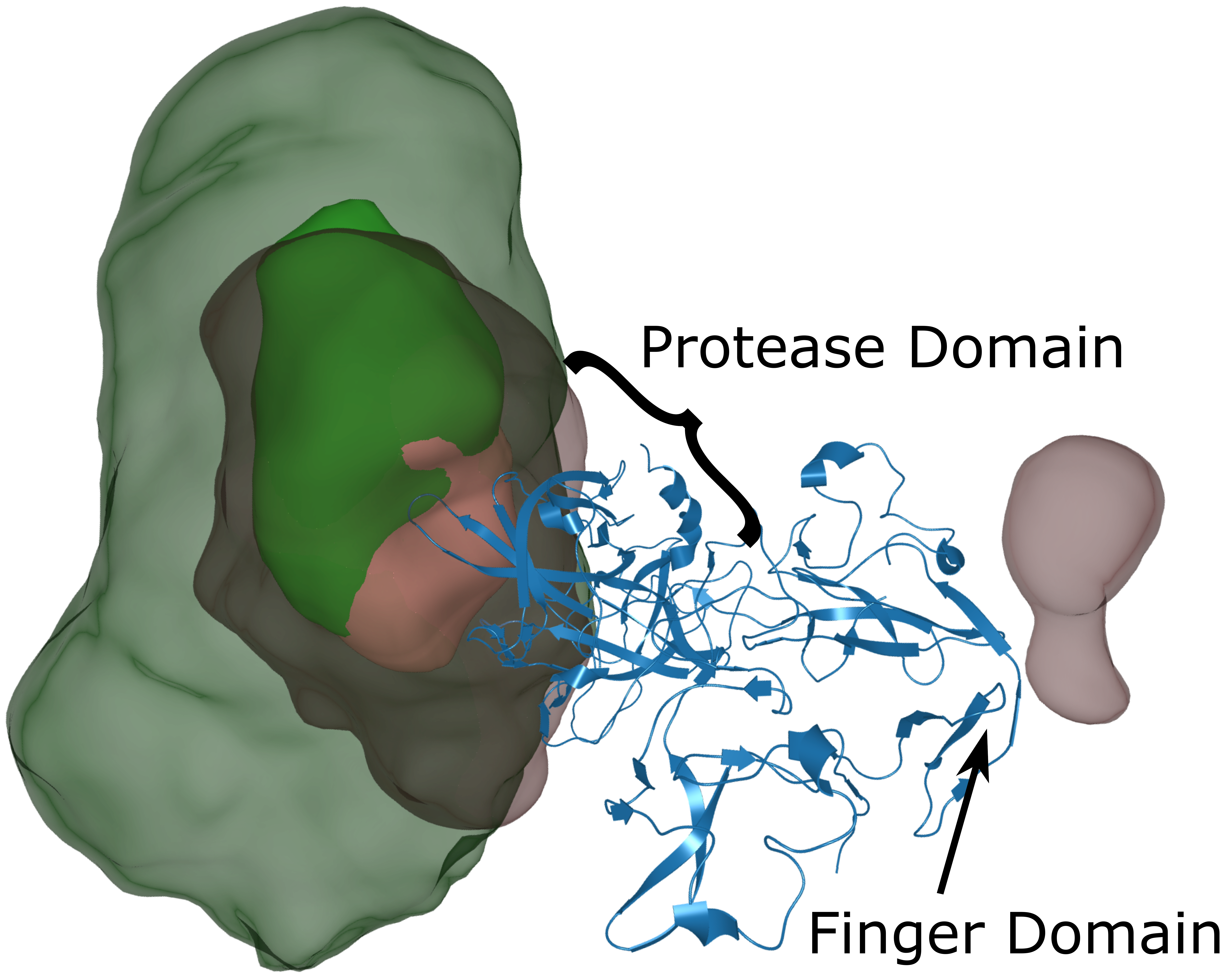}
	\caption{3D Density Overview of \CDFULL. In this case, t-PA protein (blue) was selected as the primary protein. The green and pink isosurfaces then indicate the positions of PAI-1 and LRP proteins \rev{w.r.t. to}  the primary protein. We see that while PAI-1 seems to be interacting only with the Protease domain of t-PA, there are some configurations where LRP is interacting with the Finger domain of t-PA.}
	\label{fig:3d_view}
	\vspace{-5pt}
\end{figure}

\paragraph{3D Density Overview}
To overcome the occlusion problem of ensembles of \CC{}s and still giving an overview of their spatial arrangements, we propose to use an isosurface visualization (see~Figure~\ref{fig:3d_view}).
First, we align individual \CCFULLs{} according to the selected primary protein and estimate the occurrence density of the partner proteins, i.e., all proteins in the \CCFULLs{}, which are interacting with the primary protein.
Then, for each partner protein, we estimate the density based on its atoms, using the KDE (\textit{kernel density estimation}).
Our density estimation employs an isotropic Gaussian kernel whose bandwidth models the Van der Waals protein surface.
We sample the density using a regular grid and visualize it using transparent isosurfaces showing the density levels of partner protein locations.
Both the density computation and isosurface extraction are performed on the GPU, which enables efficient evaluation of the density and real-time updates of the isosurface visualization.

\paragraph{Exploded View}
Once the proteomic experts narrow down the set of explored \CCFULLs{} to units of \CC{}s, they want to look at the 3D structure of the individual \CC{}s and their interacting \AAPFULLs{}. To solve the problem of occlusions at the contact interfaces we adopt the exploded view technique. This technique allows us to explode the proteins such that their relative spatial arrangement is preserved.
To enable the identification of contact interfaces between the proteins, we color the atoms (and corresponding surface areas) based on the 
\rev{interactions between proteins.}
Alternatively, the coloring can be changed to reflect the properties of the contact amino acids, such as the frequency of their interaction (see~Figure~\ref{fig:teaser}). 
To "explode" the \CCFULL{}, we use a force-based layout for drawing graphs.

\subsection{Filtering Interface}
\label{sec:filter_interface}
Filtering operations can be either performed based on selections of data items in the views or via dedicated range filters. Each view depicting data items with defined quantitative properties is equipped with the filtering panel that offers the range filters for the given properties. In the \rev{Property View}, for example, the filtering properties correspond to the properties defined for \CCFULLs{}, such as electrostatics and Van der Waals energy, while in the Residue Matrix it is hydrophobicty or interaction frequency of AAs and \AAPFULLs{}. As the filtering interface can be quite spatially demanding, we only show it on demand.

We provide the users with the overview of all applied filters in a separate Filter View (see Figure~\ref{fig:teaser}f). It consists of two parts. The status bar at the top of the view shows how many \PPCFULLs{} were filtered out and how many are remaining in the explored ensemble. 
Below the status bar, the list of all applied filters is shown. For each filter, \rev{we indicate its type, subject, and properties}, e.g., the type of the data item, the size of selection that was filtered, or the range of applied range filter. This way the users can see the steps they undertook during the exploration process and adjust them (\TPROV). Each filter can be temporarily disabled or removed. When a filter is disabled, all filters are re-evaluated and the data items that were filtered out by this filter will reappear in the individual views with the indication that they are affected by the disabled filter. In the Overview Graph and \rev{Property View}, the filtered portion of the ensemble will be grayed out. In the \rev{Protein View} and Residue Matrix, we did not want too loose the color mapping that is an essential part of these views. Moreover, these views show the aggregated information, and as such, their cells may be only partially affected by the filters. This happens, for example, if a cell in the Residue Matrix represents an \AAPFULL{} present in 20 \CCFULLs{}, 5 of which are affected by the disabled filter. Therefore, we chose to indicate the filtered state of the items in these views by a white cross. The full cross indicates that the item is completely filtered by the disabled filter, i.e., after re-enabling of the filter the item will disappear. A single diagonal through the item cell indicates that the cell is partially affected by the filter. This way the users can effectively study the effects of individual filters and possibly re-add the filtered items back to the ensemble. Naturally, multiple filters can be disabled at the same time and there is also the possibility to disable or remove all the filters at once.





\section{Interaction Patterns}
During the evaluation, when the domain experts used our tool, we observed several common interaction patterns. Since we think that these patterns are likely to occur in other cases of hierarchical data exploration, we comment on them in the following.

Upon loading an ensemble of protein complexes into our tool, users are presented with an abstract representation of the ensemble in the form of the Overview Graph. We noticed that after briefly checking this representation, and before proceeding with their analysis, the experts usually looked at few examples from the explored ensemble using standard molecular visualization techniques. We think that they did so to improve their mental link between the new abstract representation and the actual data in the ensemble -- enabling this was probably vital for the successful application of the abstraction.

We further noticed that even when the users were familiar with the explored data and knew what they were looking for, they still needed to switch multiple times between different levels of abstraction
before they decided on how to proceed. 
We observed that users tended to spend more time focusing on the middle levels of the hierarchy (\rev{Protein View}, 3D Density Overview, and Residue Matrix).
This was most apparent in the early stages of the exploration when the users familiarized themselves with the dataset, as well as when searching for patterns and features that they considered for filtering.
This seems to confirm that providing direct interaction, enabling the swift navigation between the individual levels of abstraction, is crucial for a successful exploration of hierarchical data.

Also the ability to relate one view to others proved to be vital. In our case, users utilized the selection operations to get more information about the selected element (e.g., selection of the \AAPFULLs{} in the \rev{Protein View}, which led to their selection in the Residue Matrix, where the users could see both AAs forming the pair; and the selection of a configuration in the \rev{Property View} to see its 3D structure). 

We also observed a back-and-forth pattern later in the process when applying filtering operations. It became clear that disabling and reverting filters was one of the most commonly used features of our tool and that it was vital for the successful drilldown process. Therefore, providing users with explicit information about how a filtering operation affects the data proved to be crucial. 


\section{Case Studies}
\label{sec:casestudy}



Here we demonstrate the use of our tool on two case studies performed by our collaborating proteomic expert. \rev{The expert had an experience with the previous version of the COZOID tool and after a short (about 30 minutes) explanation of the new views, he was ready to use the application by himself with occasional help from the authors.}

\subsection{Human t-PA Model}
In the first case study, the expert analyzed the results of the protein docking performed using HADDOCK multi-body docking server.
The explored ensemble contained predictions of spatial arrangements for the protein complex consisting of three proteins. The first protein was tissue-type Plasminogen activator (t-PA), which is the initiator of the dissolution of blood clots in humans and other species. The second protein was Plasminogen activator inhibitor (PAI-1) that blocks the function of t-PA. It is proven to bind to the catalytic site of the Protease domain of t-PA via its reaction center loop. The third molecule was a part of the lipoprotein-like receptor (LRP) complement-type repeat. It is expected to bind either to one of the basic amino acids of the PAI-1 helices or to a Finger domain of t-PA. The input data for the docking procedure were taken from the publication of human t-PA model~\cite{rathore2011first} and the crystal structure published under PDB ID 5BRR. \rev{The goal of the study was to investigate if the docking tool can predict a reasonable configuration of the proteins that would correspond to the experimental knowledge the domain expert has about the complex.} 

The HADDOCK results contained 200 possible \CCFULLs{} of the three proteins. From these, \rev{HADDOCK} selected \rev{ten} \CCFULLs{} as the best representatives of the docking run based on the similarity clustering and internal scoring function. The proteomic expert first loaded just this subset, expecting that the most viable solution would be found in these representatives. In the Overview Graph, he selected the t-PA protein and looked at the 3D Density Overview. There he saw that in most of the cases the LRP molecule was bound close to PAI-1 and Protease domain of t-PA. However, there also seemed to be some \CCFULLs{} where LRP and PAI-1 were bound to the opposite side of t-PA. This \rev{observation} was confirmed when the expert quickly looked on the 3D representations of individual \CCFULLs{}. He noticed there was precisely one \CCFULL{} where this occurred. While for LRP this binding is possible, it made no sense for PAI-1 protein. Thus this \CCFULL{} was removed from the ensemble. The expert also noticed that in cases where the PAI-1 was bound to the correct domain of t-PA, its orientation seemed to vary greatly. To see if the binding site, which is already well documented in the literature, was preserved despite the varying orientation, he selected the \PPDFULL{} of t-PA and PAI-1 for exploration in Residue Matrix. Here he selected \AAPFULL{} R299 (t-PA) and D222 (PAI-1), as these charged AAs have been proven to influence binding of the two proteins considerably, and removed the \CCFULLs{} that did not contain this pair. He was left with three \CCFULLs{}. He then switched the explored \PPDFULL{} to PAI-1 -- LRP. After looking at the Contact Zone List View, he concluded that the position of the LRP molecule was not bound correctly in any of these cases.

\begin{figure}[b]
	\centering
	\includegraphics[width=\columnwidth]{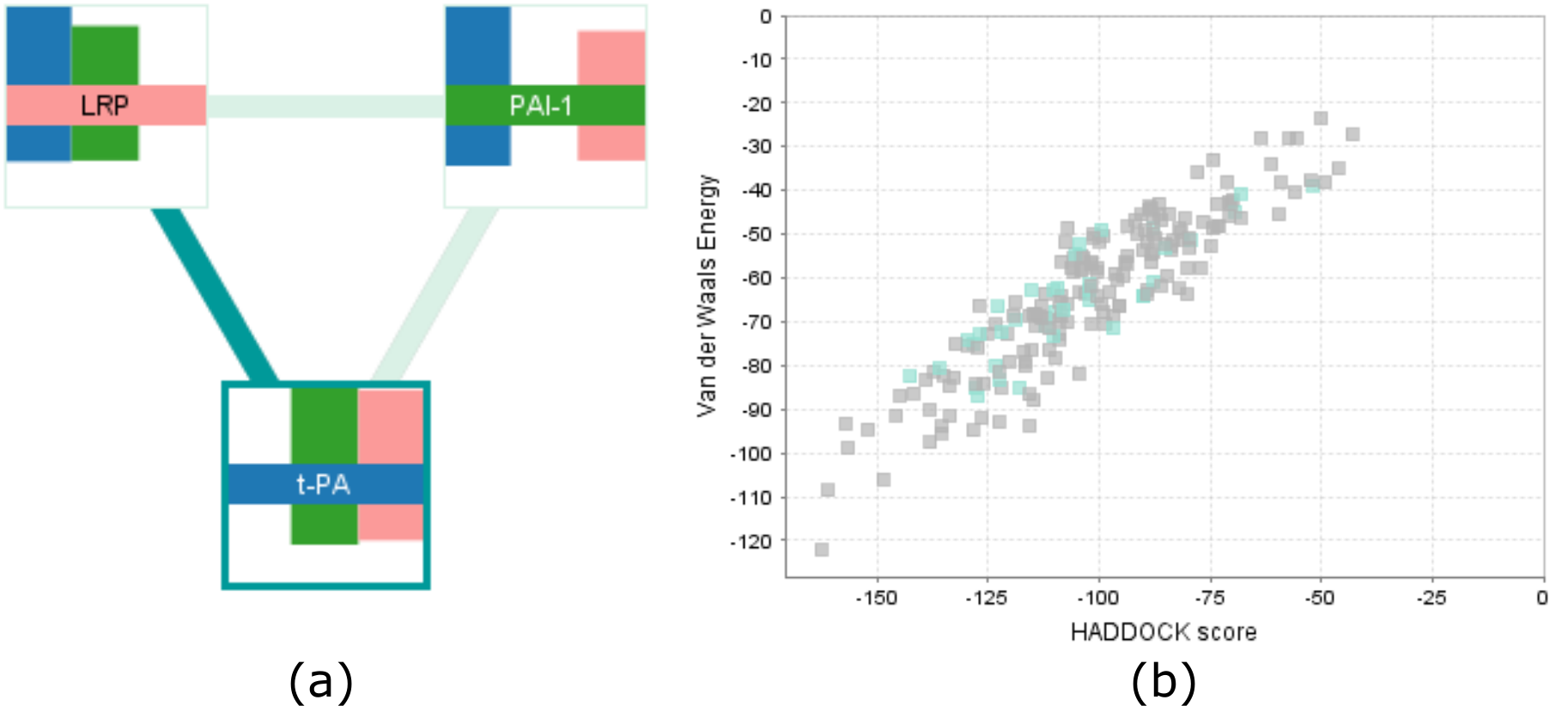}
	\caption{a) Overview Graph of the tPA -- PAI-1 -- LRP docking results. The small \revBar{height} of the colored \revBar{bars below} the core of each node indicates that the contact interfaces between the proteins are not very stable. b) \rev{Property View} showing the relationship between the HADDOCK score and Van der Waals energy. The filtered out \CCFULLs{} are depicted in gray.}
	\label{fig:casestudy-overview}
\end{figure}

Seeing that the preselected subset did not contain any viable solution, the expert then loaded a full set of 200 \CCFULLs{} into our system. The Overview Graph (Figure~\ref{fig:casestudy-overview}a) indicated that the binding sites were not very stable. Therefore, the expert again selected t-PA as the primary protein and looked at the 3D Density Overview to check the spatial distribution of the data (Figure~\ref{fig:3d_view}). The situation was similar to the previously examined subset -- both LRP and PAI-1 molecules were centered around the Protease domain of t-PA with a large variance in positioning. The wrongly bound PAI-1 was not visible in this case (the density of such solutions was below the isosurface threshold). However, a portion of the ensemble seemed to contain LRP binding to the Finger domain of t-PA.

Based on the experience with the subset of \CCFULLs{}, the expert expected that the large portion of t-PA -- PAI-1 \PPDFULL{} would be infeasible. Therefore, he selected this \PPDFULL{} and repeated the \AAPFULL{} filtering in the Residue Matrix (R299 (t-PA) -- D222 (PAI-1)). This \rev{operation} immediately removed all but 35 \CCFULLs{}.

The expert wanted to see how the remaining \CCFULLs{} performed according to different scoring functions, curious why the HADDOCK selected representatives did not contain more biochemically relevant solutions. He wanted to compare the scores of already filtered out \CCFULLs{} to the ones he was still considering. Therefore, he temporarily disabled the filtering and turned his attention to the \rev{Property View}. As he browsed through different combinations of properties, he observed that the HADDOCK score was closely related to Van der Waals energy (Figure~\ref{fig:casestudy-overview}b). However, none of the properties seemed to correspond to his own filtering in any way. So he concluded that in this instance he \rev{could not} rely on the pre-computed scores.

The expert then decided to proceed with the exploration of the LRP bindings. He knew from the 3D Density Overview that in the majority of \CCFULLs{} the LRP molecule was located close to the t-PA -- PAI-1 binding site and decided to explore this part of the dataset first. Thus he wanted to remove the \CCFULLs{} where this was not the case. As the t-PA was still selected as the primary protein, the expert switched to the \rev{Protein View} to locate the Finger domain amino acids with the LRP binding. He selected the corresponding \AAPFULLs{} and \CCFULLs{} where these \AAPFULLs{} were interacting (Figure~\ref{fig:casestudy-1D}) and then filtered out the selection. This \rev{operation} removed only three \CCFULLs{}.

\begin{figure}[h]
	\centering
	\includegraphics[width=0.7\linewidth]{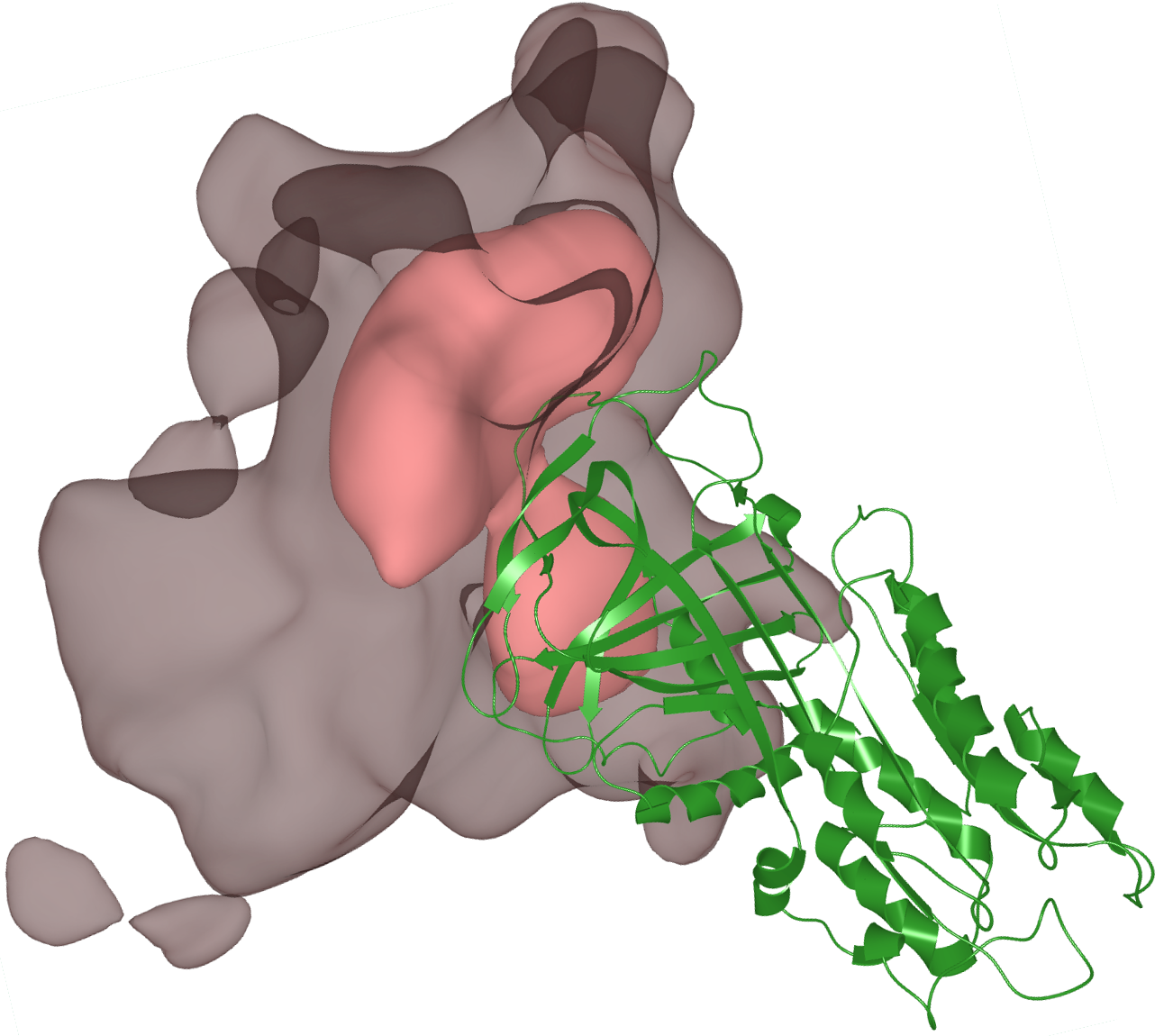}
	\caption{3D Density Overview of the subset of 35 \PPCFULLs{} of PAI-1 (green) with LRP (pink).}
	\label{fig:casestudy-densityPTA}
\end{figure}

\begin{figure}[h]
	\centering
	\includegraphics[width=0.88\linewidth]{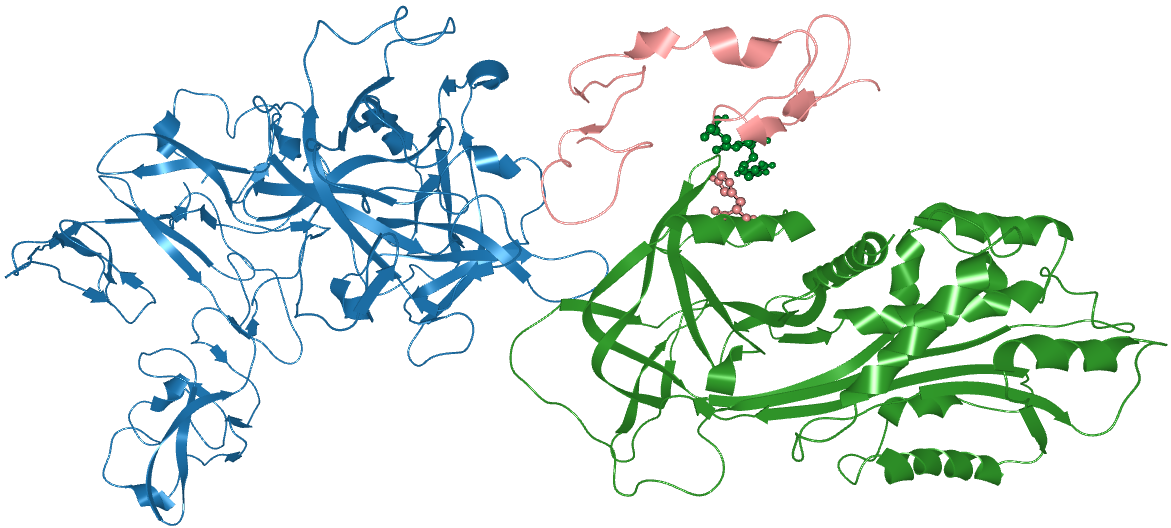}
	\caption{\CCFULL{} with LRP (pink) interacting with basic residue H261 on PAI-1 (green) helix. It can be seen that the position of LRP disrupts the interaction between t-PA (blue) and PAI-1.}
	\label{fig:casestudy-case1}
\end{figure}

The expert knew that LRP was proven to bind to a basic amino acid on the PAI-1 helix. Therefore, he wanted to find out if any of the \CCFULLs{} fulfilled this condition. He switched the primary protein to PAI-1. In the 3D Density Overview he saw that LRP molecules in the ensemble were close enough to only one helix of PAI-1 -- the helix closest to the binding site with t-PA (Figure~\ref{fig:casestudy-densityPTA}). In the \rev{Protein View}, he looked up the amino acid sequence part corresponding to this helix and its only basic amino acid, H261. He removed the \CCFULLs{} where LRP had no interaction with this amino acid and ended up with only one \CCFULL{}. However, looking at its 3D structure (Figure~\ref{fig:casestudy-case1}), he was immediately able to see that this \CCFULL{} was improbable since the location of LRP was acting as a wedge between t-PA and PAI-1.

\begin{figure}[h]
	\centering
	\includegraphics[width=\linewidth]{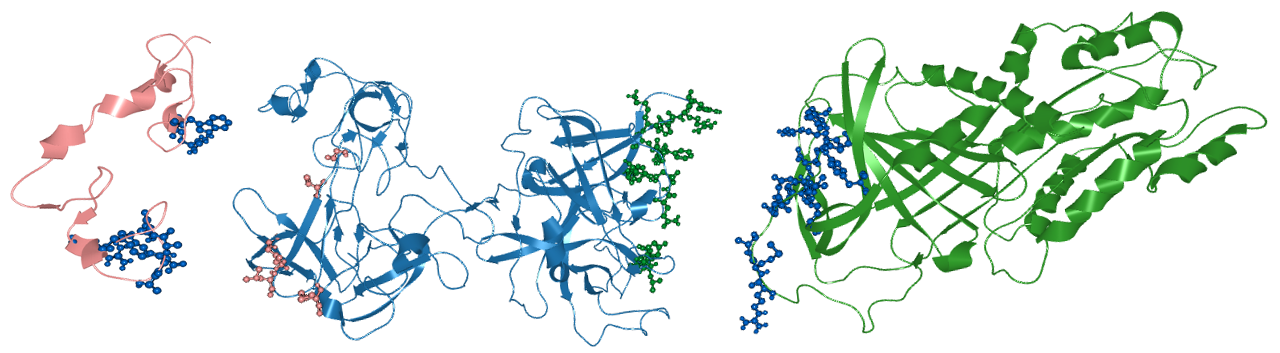}
	\caption{Exploded view of the \CCFULL{} with the best binding between t-PA (blue) and LRP (pink) showing the interacting AAs. It can be seen that unlike the case in Figure~\ref{fig:casestudy-case1}, the interaction between the t-PA and PAI-1 is preserved here.}
	\label{fig:casestudy-case2}
\end{figure}

Seeing that he was unable to find a relevant \CCFULL{} with LRP binding to the PAI-1, the expert decided to explore the smaller subset where LRP was binding to the Finger domain of t-PA. He removed all the filters except for the first one, returning to the point where he had 35 \CCFULLs{}. He then repeated the selection of \CCFULLs{} with interactions at the Finger domain of t-PA, this time removing the complementary subset, so he ended up with three \CCFULLs{}. Using the Contact Zone List View and 3D representations of individual \CCFULLs{}, he selected the one with the tightest contact with t-PA (Figure~\ref{fig:casestudy-case2}). The expert claimed that this \CCFULL{} was not expected 
but it has possible biological importance, which should be further studied. 

He further wanted to find out if there were more similar \CCFULLs{} with regard only to the binding at the Finger domain of t-PA (i.e., disregarding the t-PA -- PAI-1 interaction). He therefore temporarily disabled the filters and again selected all \CCFULLs{} with interaction at the Finger domain. This \rev{operation} also selected previously filtered \CCFULLs{}. The expert then re-added these \CCFULLs{} to the ensemble, which yielded 41 \CCFULLs{}. He then sorted the \rev{Property View} according to the similarity to the initially selected \CCFULL{}. He saw that most of the \CCFULLs{} exhibited quite a low similarity. He selected five \CCFULLs{} with the highest scores to compare in the Contact Zone List View and 3D, but concluded that the first structure was still the best among them. The Contact Zone List View showing the t-PA -- LRP \PPCFULLs{} of the best \CCFULL{} and the two most similar \CCFULLs{} is shown in Figure~\ref{fig:cozolist}. 

In this case study, the proteomic expert analyzed an ensemble of 200 \CCFULLs{}. After he checked the \CCFULLs{} preselected by the docking tool using his usual tools, he stated that he does not believe the docking tool was successful in predicting a potentially relevant \CCFULL{}. We then asked the domain expert to re-evaluate the ensemble using our solution. He again started with the analysis of the preselected \CCFULLs{}, which confirmed his initial statement. However, he then continued with the analysis of the full ensemble, where he eventually found one \CCFULL{} that he considered possibly biologically important and that should be explored further. He concluded that finding this \CCFULL{} without our solution would be virtually impossible. He would not be inclined to explore this dataset in detail, because it would take him an incredible amount of time and given the initial results, he did not think there was a chance of finding anything interesting. He stated that even if he chose to explore the ensemble in greater detail, he would easily miss the one \CCFULL{} he selected in the end, as it did not have the expected spatial arrangement. However, with our solution, he was easily able to filter out unsuitable \CCFULLs{}, focus on interesting subsets, and find their best representatives. \rev{Our solution} enabled him to explore the ensemble in a faster manner and discover new and unexpected results.

\subsection{Human Nucleosome}
To verify the scalability of our solution, we wanted to test it also on an ensemble of a larger protein complex. Therefore, we created an ensemble of prediction for human nucleosome structure (published under PDB ID 3AFA). This structure consists of eight proteins. The ensemble was prepared by taking pairwise predictions for five pairs of proteins produced by pyDock~\cite{cheng2007pydock}, which were then iteratively combined to recreate the nucleosome structure. 
In this phase, the colliding combinations were automatically discarded. 
Then, a representative sample of 500 solutions was selected, which included
a set of 35 \CCFULLs{} closest to the published crystal structure as well as randomly sampled decoys. 

We then loaded the data into our system and asked the proteomic expert who provided us with the data to identify the best solutions. The expert was already familiar with the interaction patterns in the nucleosome structure and thus he began his exploration in the Overview Graph, where he removed \PPDFULLs{} which were not supposed to interact. He then continued his analysis using the Residue Matrix, where he continued with the filtering based on the most prominent contacts (e.g., the salt bridge between units H2A and H2B), until he finished with a set of 11 configurations. He then loaded the crystal structure to our tool and used the Contact Zone List View to compare \rev{it} with the selected structures. He found out that the majority of structures was reasonably similar to the crystal structure. A snapshot from this exploration can be seen in Figure~\ref{fig:teaser}. In terms of scalability, the tool was fully interactive even for this larger ensemble. The performance of the tool only differs in the initial precomputation time, which may take several minutes, depending on the size of the ensemble. However, the precomputation time is not a limitation for the experts. 

\section{Feedback and Discussion}
\rev{We have evaluated our system with the domain experts from two different research groups.}
Here we summarize the most notable feedback \rev{and discuss also some limitations of our approach}.

The proteomic experts stated that the Overview Graph together with the 3D Density Overview provide a nice initial overview of the ensemble. They particularly liked the graph stating that it is useful especially for protein complexes consisting of a higher number of protein units, as the interaction patterns become more intricate and harder to analyze with the increasing size of the complex. 


\revBar{When designing our solution, we have considered multiple layouts for the Overview Graph. In the end, we decided to stick to the circular layout as it communicates that the visualization is abstract. The non-uniform layout could suggest that some proteins are spatially closer than the others which would be misleading as we cannot satisfactorily compute these spatial properties from the dataset.}
\rev{We have also considered different mappings (instead of the \revBar{bar} chart) to communicate the size of the contact interfaces. However, as the contact interface size is tightly connected with its consistency, we always need to depict these two values close to each other. If we would choose, for instance, the edges of the graph for this mapping, we would also need two information channels, \ie edge width and its color. We do not consider this a suitable combination, as the color interpretation might be distorted due to the varying width of the lines. 
}


Regarding the filtering interface, the users appreciated the ability to disable the filters and observe their effect as well as the possibility to revert filters and thus return to the previous points in exploration. They noted that it would be nice to have the possibility to group the applied filters and annotate them with their own notes so it would be easier to navigate in the filter list and they could disable or remove the groups together. We plan to \rev{add the support for filter grouping} in the near future. However, considering that the results of filtering depend on the order of filters (in particular the position of the \textit{add} operation in the filter queue), the grouping needs to be restricted to preserve the initial order of the filters.

Related to this was the request to save the exploration state for future analysis. Our framework already partially supports this via the ability to store workspaces, which includes storing of selected configurations, and we are currently extending it to support also the storage of filter states.

The proteomic experts also listed several properties that could be added to our tool, such as amino acid conservation or the type of bond in the \AAPFULLs{}. Our solution can be easily extended with these. On a similar note, we would also like to provide the users with the support for importing the precomputed \CCFULL{} scores from additional docking tools. According to domain experts, it is more reliable and informative to use the scores computed during docking simulation than re-scoring them afterwards. 

Overall, the proteomic experts appreciated the ability to navigate through different levels of data hierarchy and use them to filter and select the \CCFULLs{}. They stated that our tool enabled them to explore their ensemble in a manner that was not previously possible. It considerably decreased the time spent on the analysis process and enabled them to identify solutions they would otherwise not be able to find.

\rev{We would also like to discuss the limitations and scalability issues of our application. The Overview Graph is limited by the number of proteins in a studied protein complex. As mentioned, this number in known complexes rarely exceeds 10. However, we cannot assure that this number will not be higher in the future. In such cases, the Overview Graph may become too cluttered to be used effectively. It is also limited by the number of colors that are currently used to distinguish between individual proteins, as it is not recommended to use more than 7--12 colors for such an application~\cite{harrower2003colorbrewer}.} 

\rev{Regarding the performance scalability, the most demanding step is the preprocessing of the data, which includes the computation of the contact interfaces, various properties, and initialization of the hierarchical data structures. The performance in this step has an asymptotic complexity $\mathcal{O}(m^2*n)$ where $m$ is the number of proteins in the complex and $n$ is the number of individual \CCFULLs{}. We have tested the application on the dataset containing 500 \CCFULLs{} of a protein complex consisting of 8 proteins. Here the initialization step took several minutes. However, after this the application was fully interactive in real time.}

\section{Conclusion and Future Work}
We have presented a novel approach to the analysis of large ensembles of multi-body protein complexes. This approach solves the problem of analyzing hundreds of possible spatial configurations of protein complexes produced by a docking tool and selecting suitable representatives that can be further explored in the lab. Our approach builds on the hierarchical structure of protein complex ensembles which we use to facilitate the exploration of the data at multiple levels of the hierarchy.

We have evaluated the usefulness of our approach in a case study performed by a proteomic expert. In the case study, the expert was able to use our tool and find an important configuration of the explored protein complex in the ensemble, which he initially marked as uninteresting based on preselected samples. This was a previously unattainable task, as the domain experts lacked proper tools to explore the entire ensemble in detail.  
In the future, we would like to extend our solution with the possibility to automatically cluster the results based on the similarity of the contact interfaces and spatial arrangement, and provide means for comparison of groups of \CCFULLs.

Another topic that came up during the design process of our system was to search for a suitable scoring function. As we have stated in Section~\ref{sec:data_tasks}, different tools use different scoring functions that are based on a combination of energetic, physical, and evolutionary properties and there is no single best option. Thus finding a suitable scoring function is a challenging task dealing with multi-objective optimization and it could benefit greatly from our visual guidance.

\acknowledgments{
We would like to thank the domain experts Jan Mi\v{c}an and Jan Pale\v{c}ek for providing us with the datasets, valuable feedback, and conducting the case study. 
This work was supported by the Czech Science Foundation international project GC18-18647J and the Internal Masaryk University grant (MU/0822/2015). Parts of this work were done in the context of the VIDI project, which is supported by Bergens Forskningsstiftelse, the Mohn Medical Imaging and Visualization Center (MMIV), and the University of Bergen.
}

\bibliographystyle{template/abbrv-doi-hyperref}

\bibliography{refs.bib}
\end{document}